\documentclass{aastex}
\usepackage{emulateapj5}

\shorttitle{Stellar kinematics in Leo\,II}
\shortauthors{Koch et al.}
\slugcomment{Accepted for publication in the AJ, April 25, 2007}
\def\kms{\,km\,s$^{-1}$}
\def\p1{Paper\,I}

\begin{document}

\title{Stellar kinematics in the remote Leo\,II dwarf spheroidal galaxy -- \\
Another brick in the wall\altaffilmark{1}}

\author{Andreas Koch\altaffilmark{2,3}, 
        Jan T.~Kleyna\altaffilmark{4}, 
	Mark I.~Wilkinson\altaffilmark{5,6}, 
        Eva K.~Grebel\altaffilmark{2,7}, \\
	Gerard F.~Gilmore\altaffilmark{5}, 
        N.~Wyn Evans\altaffilmark{5},
	Rosemary F.~G.~Wyse\altaffilmark{8}, 
	Daniel R.~Harbeck\altaffilmark{9}}
\email{akoch@astro.ucla.edu}

\altaffiltext{1}{Based on observations collected at the European Southern 
Observatory at Paranal, Chile; proposal 171.B-0520(A), and on observations 
made through the Isaac Newton Groups' Wide Field Camera Survey Programme 
with the  Isaac Newton Telescope operated on the island of La Palma by the 
Isaac Newton Group in the Spanish Observatorio del Roque de los Muchachos 
of the Instituto de Astrofisica de Canarias.}
\altaffiltext{2}{Department of Physics and Astronomy, Astronomical Institute of the University of Basel,
Binningen, Switzerland}
\altaffiltext{3}{UCLA, Department of Physics and Astronomy, Los Angeles, CA, USA}
\altaffiltext{4}{Institute for Astronomy, University of Hawaii, Honolulu, HI, USA}
\altaffiltext{5}{Institute of Astronomy, Cambridge University, Cambridge, UK}
\altaffiltext{6}{Department of Physics and Astronomy, University of Leicester, Leicester, UK}
\altaffiltext{7}{Astronomisches Rechen-Institut, Zentrum f\"ur Astronomie Heidelberg,
University of Heidelberg,  Heidelberg, Germany}
\altaffiltext{8}{The John Hopkins University, Baltimore, MD, USA}
\altaffiltext{9}{Department of Astronomy, University of Wisconsin, Madison, WI, USA}

\begin{abstract}
We present the projected velocity dispersion profile for the remote (d=233\,kpc)
Galactic dwarf spheroidal (dSph) galaxy Leo\,II, based on 171 discrete
stellar radial velocities that were obtained from medium-resolution
spectroscopy using the FLAMES/GIRAFFE spectrograph at the European
Southern Observatory, Chile. The dispersion profile of those stars with good 
membership probabilities is essentially
flat with an amplitude of 6.6$\pm$0.7\kms\ over the full radial extent
of our data, which probe to the stellar boundary of this galaxy. We
find no evidence of any significant apparent rotation or velocity
asymmetry which suggests that tidal effects cannot be invoked to
explain Leo\,II's properties.  From basic mass modeling, employing Jeans'
equation, we derive a mass out to the limiting radius of ($2.7\pm
0.5)\times10^7\,M_{\odot}$ and a global mass to light ratio of 27--45 in
solar units, depending on the adopted total luminosity.  A cored halo
profile and a mild amount of tangential velocity anisotropy is
found to account well for Leo\,II's observed kinematics, although we
cannot exclude the possibility of a cusped halo with radially varying
velocity anisotropy. 
All in all, this galaxy exhibits dark 
matter properties which appear to be concordant with the other dSph 
satellites of the Milky Way, namely a halo mass profile which is 
consistent with a central core and a total mass which is similar to 
the common mass scale seen in other dSphs. 
\end{abstract}


\section{Introduction}

In the two decades since the seminal work of Aaronson (1983) on the
velocity dispersion of the Draco dwarf spheroidal galaxy (dSph), it
has become observationally well-established that all of these lowest
luminosity galaxies exhibit large overall velocity dispersions (of the
order of 10\kms). Further, their radial velocity dispersion profiles
are generally flat to very large radii (Kleyna et al. 2002, 2004; Koch
et al. 2007a; Mu\~noz et al. 2006; Walker et al. 2006; though see
Wilkinson et al. 2004 for the case of Draco and Ursa Minor). Coupled
with their relatively large characteristic radii (of the order of
hundreds of parsecs; see Fig.~1 in Gilmore et al. 2007) and low,
star-cluster-like luminosities (Mateo 1998), these observations have
led to the conclusion that these systems may be dominated by large
amounts of dark matter on all spatial scales (see Mateo 1998; Gilmore
et al. 2007 for recent reviews). Their estimated mass-to-light (M/L)
ratios are often large, with values as high as several hundreds being
reported in the literature\footnote{Throughout this paper, we will
state all values of M/L in terms of $V$-band luminosity and solar
units.}. The recent addition of eight new dSph candidates to the
census of Milky Way satellites (Belokurov et al. 2006,2007; Irwin et
al. 2007; Zucker et al. 2006a,b) has highlighted the bi-modal nature
of the distribution of sizes for low-luminosity stellar systems. All
known star clusters exhibit half-light radii smaller than 30pc while
no dSph has a stellar core radius smaller than about 120pc (Belokurov et
al. 2007; Gilmore et al. 2007). Given the apparent absence of dark
matter in star clusters, this bi-modality adds circumstantial weight
to the suggestion that dSphs are the smallest stellar systems to
contain dynamically significant quantities of dark matter, making them
particularly valuable for studies of its physical properties.

On the other hand, the above M/L estimates presuppose that the dSphs
are in dynamical equilibrium, an assumption which is directly
supported in, for example, the case of the Draco dSph by lack of any
significant depth extent along our line of sight (e.g., Klessen et
al. 2003).  It has been argued, though, that the flat dispersion
profiles and high apparent M/L ratios could be the results of tidal
sculpting of these galaxies (e.g., Read et al. 2006a, 2006b, Westfall
et al. 2006). While there are a few examples of close dSphs whose
shape or kinematics appear to be affected by tides (e.g., 
Ursa Major\,II [Zucker et al. 2006];
Ursa Minor [Palma et al. 2003]), the outer dSph satellites might be
expected to be relatively unscathed by tidal perturbations.  Galactic
tides, if acting, would predominantly unsettle the outermost regions
near the stellar boundary of these systems.  Moreover, the Draco dSph,
though being one of the close Galactic satellites, does not show any
indication of structural perturbations (Odenkirchen et al. 2001;
Segall et al. 2007).  Coupled with the negligible ratio of ordered
stellar rotational velocity (as a manifestation of tides; see Section 3 
below) to the random motion in dSphs, the
observations suggest that these galaxies are in fact the most dark
matter dominated systems found to date. It is nevertheless valuable to
compare the observed properties of dSphs with alternative models, and
scenarios in which the dSphs are the remnants of tidally disrupted,
dark-matter-free stellar systems have been investigated by a number of
authors (e.g. Kroupa 1997; Kuhn \& Miller 1989; Klessen \& Zhao 2002).

Initially, information about the dSphs' dark matter content was mainly
gleaned from estimates of their central velocity dispersion.
Fortunately, in recent years wide field surveys have provided the
opportunity to analyze stellar data covering the entire spatial extent
of the dSphs and to investigate stellar kinematics even beyond
their formal tidal limits as derived from King profile fits to their
stellar density distributions (usually some tens of arc minutes; e.g., 
Odenkirchen et al. 2001; Kleyna et al. 2003, 2004; Tolstoy et al. 2004; Chapman et al. 2005;
Mateo 2005; Mu\~noz et al. 2006; Walker et al. 2006; Wilkinson et
al. 2006 and references therein; Koch et al. 2007a). Such studies are
essential to reliably distinguish between the dark matter dominated
and tidal disturbance scenarios.

The mass-density profiles of low-mass galaxies provide valuable
information about the spatial distribution of dark matter on small
scales. It has been suggested that halo density profiles, in
particular the differentiation between cored and cusped halos, can
provide an insight into the actual nature of cold dark matter (CDM;
Read \& Gilmore 2005; Colafrancesco et al. 2007; Gilmore et
al. 2007). Numerical CDM simulations with high spatial resolution
predict dark matter halos that exhibit a central density cusp,
$\rho(r)\propto r^{-\alpha}$ with $\alpha$=1--1.5 (e.g., Navarro,
Frenk \& White 1995; hereafter NFW; {Diemand et al. 2005}). 
By
contrast, observations of the kinematics of massive galaxies are fully
consistent with flat density cores (e.g., Gentile et al. 2004;
Spekkens et al. 2005), which also appear to account for most of the
available data for dSphs (\L okas 2002; Kleyna et al. 2003; Read \&
Gilmore 2005; Goerdt et al. 2006; S\'anchez-Salcedo et al. 2006;
Strigari et al. 2006; Wilkinson et al. 2006; Koch et al. 2007a, {Wu 2007}). The
presence of central cores in the dark matter haloes of dSphs has
important implications for the properties of dark matter as it may
imply a maximum phase space density for dark matter which can be used
to discriminate between competing dark matter candidates (CDM, Warm
Dark Matter, self-interacting dark matter, etc.). One should bear in
mind, however, that in the case of dSphs such inferences are subject
to low-number statistics and uncertainties in the models and do not
take account of the dynamical evolution of the dwarf galaxies'
progenitor systems (Grebel et al. 2003).

In the light of all the above, the remote dSph Leo\,II provides an
interesting test bench to investigate the dark matter content of Local
Group dSphs. At its present-day Galactocentric distance of 233\,kpc
(Bellazzini et al.  2005) it is unlikely to have experienced severe
Galactic tides unless it is on a very eccentric orbit that brings it
close to the Milky Way.  Previous kinematic data for this satellite
were assembled in the study of Vogt et al. (1995; hereafter V95) for
31 stars within the core radius at $2.9\arcmin$ (Irwin \&
Hatzidimitriou 1995).  From these data, V95 found a M/L of 7 and
concluded that Leo\,II must be embedded within a massive dark matter halo,
although its dark matter properties are not extreme compared to those
of other dSphs.  On the other hand, Coleman et al. (2007), using photometry 
from the Sloan Digital Sky Survey, 
estimate a M/L of the order of 100, but predicate this
value on the assumption that Leo\,II is purely dark matter dominated.

In this Paper, we shall investigate the dynamics of this particular
galaxy by analyzing data for a large number of targets that reach out
to the nominal outer optical radius at $\sim9\arcmin$ (Irwin \& Hatzidimitriou
1995; Coleman et al. 2007) thus enabling us to perform a full
kinematical analysis of Leo\,II and to derive detailed information on its
density distribution, its global M/L ratio and thus to answer the
question of how well-behaved or extreme this galaxy is in terms of its
dark matter content.
 
This paper is organized as follows: our data, their reduction and the
derivation of individual velocities are described in \textsection
2. We investigate the extent of Leo\,II's apparent rotation in
\textsection 3 and determine the galaxy's radial velocity dispersion
profile in \textsection 4. These results allow us to derive mass and
density profiles in \textsection 5, while \textsection 6 is dedicated
to the question of whether Leo\,II's density profile shows a cusp or a
core, as well as the importance of velocity anisotropy.  Finally, we
summarize our findings in \textsection 7.

\section{Observations and reduction}

In the framework of the ESO Large Programme 171.B-0520(A) (principal
investigator: G.~F. Gilmore), which aims at elucidating the kinematic
and chemical characteristics of Galactic dSphs, five fields in Leo\,II
were observed using the FLAMES multi-object facility at ESO's Very
Large Telescope (Pasquini et al. 2002). Within the same programme, we
also analysed the Carina dSph, which is described in detail in Koch et
al. (2006).  In a first paper concerning Leo\,II that made use of these
data we presented this galaxy's spectroscopic metallicity and age
distributions, which were derived using the near-infrared calcium
triplet (CaT) calibration method (Koch et al. 2007b, hereafter \p1).
For details on our observing strategy, the data reduction and
calibration techniques, we refer the reader to \p1. In this section we
briefly summarize the main steps taken.
\subsection{Target selection and acquisition}
The wide field of view of the FLAMES instrument, with a diameter of
25$\arcmin$, enables one to cover the whole projected area of the
Leo\,II dSph with one single pointing, since its nominal tidal radius
lies at 8$\farcm$7 (Irwin \& Hatzidimitriou 1995).  However, in order
to achieve good sampling out to large radii and to yield more
information about the stellar content near and beyond the tidal
radius, we observed five different (overlapping) fields in several
instrument configurations, typically offset by 5$\arcmin$ with respect
to each other (see Tables 1 and 2 in \p1).

Our targets were drawn from photometry that was obtained by the
Cambridge Astronomical Survey Unit\footnote{see
http://www.ast.cam.ac.uk/$\sim$wfcsur} (CASU; Irwin \& Lewis 2001) at
the 2.5\,m Isaac Newton Telescope (INT) on La Palma, Spain.  From this
data set we selected red giant candidates with magnitudes ranging from
the tip of the red giant branch (RGB) at V$\sim$18.5\,mag down to
2.5\,mag below the RGB tip, reaching as deep as 21\,mag in apparent
$V$-band magnitude.
The total selected number of red giant candidates we could thus
observe was 200.  Their positions on the sky are depicted in Fig.~1.

The observations themselves were carried out using the identical
strategy and instrumental setups as described in \p1, i.e., we used
the FLAMES instrument in combination with the GIRAFFE multifibre
spectrograph in ``low-resolution''-mode centered around the
near-infrared CaT at 8550\AA. {This set up provides a full wavelength 
coverage of 8206--9400\AA\ and a resolving power of $R\sim 6500$ 
per resolution element.} 
Although we aimed at exposing each
configuration for 6\,hrs to reach a minimum nominal signal-to-noise
(S/N) ratio of 20\,{pixel$^{-1}$} at our spectral resolution of $\sim 6500$, the
major part of the nights was hampered by bad sky
conditions. Consequently, the median S/N achieved after processing the
spectra is 12\,{pixel$^{-1}$, as defined from the variance in  continuum bandpasses 
surrounding  CaT region, from which we will determine our velocities in Sect.~2.3.}.

\subsection{Data reduction}

Details of the reduction process are given in \p1. In summary, we used
version 1.09 of the FLAMES data reduction system, girbldrs, and the
associated pipeline version 1.05 (Blecha et al.\ 2000).  After
standard bias correction and flatfielding, the spectra were extracted
by summing the pixels along a slit of width 1\,pixel.  The final
rebinning to the linear wavelength regime was done using Th-Ar
calibration spectra taken during the daytime.

Sky subtraction was facilitated by the allocation of about 20 fibers
per configuration to blank sky. 
Because the pipeline did not perform reliable sky
subtraction at the time the reduction was performed, custom reduction
software was written to subtract sky background.  First, the sky
fibers were median-combined to produce an average sky for each
field. Next, the sky was subtracted from each spectrum by modeling the
observed spectrum as a second order polynomial plus the average sky,
and minimizing the integral over wavelength of the absolute residual.
The regions of the spectrum containing the lines of the CaT
were excluded from the residual calculation because they are not
expected to conform to our simple model of the spectrum.  We estimate  
the
accuracy of the final sky subtraction to be on the order of $\sim3\%$, 
defined as the 1$\sigma$-dispersion of each 
sky-subtracted spectrum divided by the median of the sky
spectrum.

Our data set was then completed by the co-addition of the
dispersion-corrected and sky-subtracted science frames, weighted by
the exposures' individual S/N, and subsequent rectification of the
continuum.
\subsection{Radial velocities and Membership estimates}
In order to separate Leo\,II's RGB stars from Galactic foreground
stars (Wyse et al. 2006), we determined the individual radial
velocities of each target star by means of cross-correlation of the
three calcium triplet lines against synthetic Gaussian template
spectra using IRAF's\footnote{IRAF is distributed by the National
Optical Astronomy Observatories, which are operated by the Association
of Universities for Research in Astronomy, Inc., under cooperative
agreement with the National Science Foundation.}  FXCOR task (see
e.g., Kleyna et al. 2004).  The templates were synthesized adopting
representative equivalent widths of the CaT in red giants. The typical
median velocity error achieved in this way is 2.4\,km\,s$^{-1}$.
Taking into account the possibility that the errors returned by FXCOR,
which are based on the Tonry-Davis R-value (Tonry \& Davis 1979), may
in fact be under- or overestimated (e.g., Mateo et al. 1998; Kleyna et
al. 2002), we {pursued two different tests of our achieved accuracy. 
First, we determined 
a multiplicative constant to the formal velocity
uncertainties by  deriving } radial velocities from
the first two of the Ca lines at $\lambda\lambda$\,8498, 8452\,\AA\
and separately from the third line at 8662\,\AA.  The constant was
then obtained by requiring a reduced $\chi^2$ of unity between these
measurements. This procedure resulted in a re-scaling of the FXCOR
velocity errors by a factor of 0.6.  
{A second estimate of our velocity errors will be 
given in Sect.~2.4 by comparing our individual measurements 
to the published values from V95.} 
A plot of velocity errors versus
apparent magnitude and {S/N} is shown in Fig.~2. 

Given Leo\,II's systemic velocity of $\sim76$\kms\ (V95), our velocity
data set may contain a number of Galactic foreground stars along the
galaxy's line of sight. However, owing to our selection of target
stars whose colors and luminosities are consistent with their being
members of Leo\,II's red giant branch and due to the much lower Galactic
field star density, Leo\,II's velocity peak clearly stands out against
the Galactic contribution (see Fig.~3; and Fig.~4 in \p1).  From the
Besan\c con synthetic Galactic model (Robin et al. 2003) we would
expect no more than 4 Galactic foreground stars within the our
color-magnitude selection box and with velocities between 40 and
120\kms, thus coinciding with
Leo\,II's intrinsic population (see also \p1).

An initial fit of a Gaussian velocity peak to the Leo\,II data yields
a mean radial velocity and velocity dispersion of 78.7\kms\ and
7.6\kms. Our sample contains 23 apparent radial velocity non-members,
which deviate by more than 5\,$\sigma$ from this initial fit and thus
were rejected from further analyses.  The mean heliocentric velocity
and line-of-sight velocity dispersion were finally determined via an
iterative error-weighted maximum-likelihood fit assuming a Gaussian
velocity distribution (e.g., Kleyna et al. 2002; Koch et al. 2007a).
From this we find Leo\,II's mean radial velocity to be
(79.1$\,\pm\,$0.6)\kms\ with a global velocity dispersion of
(6.6$\,\pm\,$0.7)\kms. These values are in good agreement with the
measurements of V95, who determined a mean velocity of
(76.0\,$\pm$\,1.3)\kms\ and a central velocity dispersion of
(6.7$\,\pm\,$1.1)\,km\,s$^{-1}$ based on 31 high-resolution spectra of
stars within the core radius at 2$\farcm$9 (Irwin \& Hatzidimitriou
1995).  If one adopts a conservative $2\sigma$-cut in the velocity
distribution in order to define membership of Leo\,II, 158 red giants are
included in our sample.  Relaxing the criterion to a $\pm 3\sigma$-cut
yields 171 radial velocity member candidates.  Given the low
likelihood of interlopers in our sample, we will apply the
$3\sigma$-cut throughout the following analyses.

One noteworthy feature in the velocity histogram (Fig.~3, right panel)
is the occurrence of an excess of stars at $\sim$85\kms, whose
velocities are approximately 0.9\,$\sigma$ higher than the sample
mean. This peak deviates by 1.8\,$\sigma$ (taken as $\sqrt{N}$) from
the best-fit curve.  We note, however,
that there is no apparent spatial correlation of the targets in this
velocity range, nor is there any particular separation in color-magnitude 
space. In order to further investigate the possibility that
we might be seeing an indication of a separate cold feature in our
data, we ran an extensive number of Monte Carlo tests, where we
generated normally distributed velocity samples centered on the
observed mean velocity of Leo\,II assuming the previously derived
dispersion, and which were additionally varied by the measurement
errors. It turns out that $\sim$30\% of the random data sets exhibit
peaks in any one bin that deviate by as much as the observed 1.8\,$\sigma$ so that
the observed excess of stars is not significant above the
1\,$\sigma$-level.
Therefore we conclude that the potential second peak in Fig.~3 is most
likely not a real velocity feature, but may rather be due to
statistical fluctuations.
\subsection{Comparison with other data}
Out of the 31 stars observed by V95, 28 coincide with targets in our
sample.  Since their data are based on high-resolution spectroscopy
obtained with the HIRES spectrograph at the Keck telescope, it is
worthwhile to employ the stars in common between the two data sets to
assess the quality and consistency of our own measurements. Fig.~4
shows a comparison of the radial velocities of these 28 stars as
measured in this work to those from V95. An error-weighted linear
least squares fit to these data yields the relation $v_{\rm
\,V95}\,=\,(0.98\,\pm\,0.10)\,v_{\rm
\,This\ work}\,+\,(0.5\,\pm\,7.6)$ with a r.m.s. scatter of 4.5\kms.
{Although the latter value yields an estimate of the expected 
uncertainty in the velocities, it is clear that the errors of  both studies and 
the individual S/N ratios have to be considered before one employs a simple re-scaling 
of our values to match the quoted r.m.s.\footnote{We note that the HIRES data of V95, 
despite their considerably higher spectral resolution than our data, are also of low S/N, 
which will contribute a non-negligible fraction to the observed scatter in Fig.~4. }. Hence, we 
determined scale factors $A$ in three bins of S/N (see right panel Fig.~4) such as 
to yield an agreement between our mean velocity errors and the scatter in terms of 
$\chi^2 = \sum  \frac{(v_{\rm This\,\,work} - v_{\rm V95})^2}{A^2\,\sigma_{\rm This\,\,work}^2 + \sigma_{\rm V95}^2} $, 
where the sum is over the stars in common. As a result, we rescaled our final errors by 
1.46 (S/N$\le$20), 1.31 ($20\le$S/N$\le 40$) and 1.18 (S/N$>$40). This leads to the uncertainties 
finally presented in Table~1.}

Recently, Bosler et al. (2007) have performed a low-resolution study
of the metallicity distribution of Leo\,II. However, their work did not
aim at a kinematical analysis of this galaxy and yielded a velocity
resolution of 54\kms\ per resolution element (at the region of the
CaT) and a median measurement uncertainty of 8.5\kms.  Hence, and
since these authors did not publish individual error estimates of
their measurements, we did not include the 37 stars that our
observations have in common with their set of 74 red giants in the
comparison in Fig.~4.  Nonetheless, an analogous fit to that described
above above yields $v_{\rm \,B07}\,=\,(0.93\,\pm\,0.22)\,v_{\rm
\,This\ work}\,+\,(12.2\,\pm\,17.5)$ and a r.m.s. scatter of 10.0\kms.

In the light of the respective uncertainties associated with each of
the published data sets, there is no significant systematic deviation
present in our measurements relative to those from V95 and the results
agree to within the errors.  On the other hand, there is an indication
of a systematic zero point offset of the velocities of Bosler et
al. (2007). However, this offset is within the uncertainties and can
be attributed to the lower resolution of the latter data set.
\section{Tests for apparent rotation}
It is often suggested that the apparent rotation of a system when
viewed in projection can be considered as characteristic of tidal
perturbation. N-body simulations, (e.g., Oh et al. 1995) indicate that
the action of the tidal field of the Galaxy on a dSph leads to
velocity gradients in the outer regions of the satellite. When viewed
in projection, these dynamical effects will lead to a systematic change
in the mean velocity along the dSphs's major axis, thus mimicking
rotation (Piatek \& Pryor 1995; Oh et al. 1995; Johnston et al. 1995;
Mateo et al. 1998; Read et al. 2006a). Hence, the most efficient way
of gathering evidence whether a particular dSph is being affected by
tides sufficiently strongly that its kinematics are being modified is
to look for any sign of apparent rotation in its outer regions.  To
date, all but one of the dSphs of the Local Group show no significant
velocity gradients. The sole exception is Carina, which exhibits some
evidence of a difference in the mean velocity at either end of the
major axis, for stars beyond the nominal tidal radius (Mu\~noz et
al. 2006).

With its present-day Galactocentric distance of (233$\pm$15)\,kpc
(Bellazzini et al. 2005), Leo\,II is not expected to have been recently
affected by tides, and will only have been perturbed in the past if
its orbit brought it into much closer proximity to the Galaxy.  This
is consistent with its observed regular structure, as pointed out by
Coleman et al. (2007).  However, since V95 argue that
Leo\,II may not have an unusually high dark matter content compared to the
inner Galactic satellites, it cannot be entirely ruled out that it may
have been tidally influenced during its evolution despite the presence
of a (possibly) considerable amount of dark matter. 

In order to investigate the question of tides, we implemented two
tests for a significant indication of rotation in the Leo\,II velocities.
First of all, in Fig.~5 we plot the run of radial velocity as a
function of major axis distance. In addition, mean radial velocities
have been determined in radial bins whose widths were chosen so as to
maintain a constant number of stars per bin (lower panel of Fig.~5).
If tides had in fact significantly perturbed the outer regions of Leo\,II,
one would expect to see a radial gradient in the velocities. In
particular, potential tidal distortions of the galaxy's outskirts,
e.g., reflected in tidal tails, could lead to distinct east-west
asymmetries in the velocity distributions and one would expect
excesses of high and low velocity stars (w.r.t. the systemic mean)
near the outer boundary on opposite sides of the galaxy.
It is worth noticing that although 13 of our 200 targeted stars lie
outside of the King-tidal radius of 8.7$\arcmin$, only two of these
lie within 5$\sigma$ of the systemic velocity.
Recent evidence suggests that Leo\,II may in fact be slightly more
extended, with a nominal tidal radius of 9.2$\arcmin$ (Coleman et
al. 2007), but as these values rely strongly on the concept
of fitting a particular functional form to surface photometry, such
differences remain insignificant and will not be pursued any further
in the present work (see Koch et al. 2007a for a detailed discussion).
In order to ascertain that we have not missed any potential gradient
or high- or low-velocity feature in our data due to our selection
criteria, we repeated the averaging in each radial bin after applying
several cuts, i.e., at 3\,$\sigma$, 5\,$\sigma$, and 10\,$\sigma$ of
the sample mean.  For each cut, the data were rebinned 
to assure a constant number of stars per bin. 
However, as Fig.~5 implies, there is no significant
radial gradient discernible in the kinematics of Leo\,II for any of the
velocity cuts.  An error weighted least squares fit to our velocity
data yields a slope of (1.2$\pm$0.9)\kms\,kpc$^{-1}$.  Further, those stars with 
higher and lower velocities than the systemic mean of Leo\,II are equally distributed
across the galaxy and there is no apparent indication of any localized
kinematic excess.

To investigate the question of the extent to which any apparent
rotational signal in terms of velocity gradients can be excluded, we
proceeded a step further and computed the mean radial velocity
difference of stars on either side of bisecting lines passing through
each individual target. In the presence of a distinct apparent
galactic rotation, such a plot should exhibit a clear sinusoidal
signal (e.g., Walker et al. 2006; Koch et al. 2007a).  These velocity
differences are shown in Fig.~6 versus the position angle of the
respective bisectors.
In fact, there is a peak with an amplitude of $\sim$2\kms\ seen in
our data, which is of the same order of magnitude as the radial
gradient discussed above.  Moreover, the maximum of this distribution
occurs at a position angle of 16.5$\degr\pm2.4\degr$, where the error
was obtained through bootstrap resampling. In the light of the large
uncertainty of Irwin \& Hatzidimtriou's (1995) quoted systemic
position angle, i.e., (12$\degr\pm10\degr$), we cannot reject a
coincidence between the apparent rotational signal in our velocity
data and Leo\,II's minor axis.
To assess the actual reality of such a kinematic gradient, we
performed 10$^4$ Monte Carlo runs, in which we generated random
samples of velocities  at the fixed sky positions
of our targets. The velocities for these tests were drawn from a
normal distribution, assuming Leo\,II's mean radial velocity and velocity
dispersion, and additionally allowing for a variation by the observed measurement
errors.
By means of these simulations we find that 87\% of the random samples
exhibit maximum velocity differences larger than our observed
value {(Fig.~6, bottom panel)}. Hence, the apparent rotational signal we detect is only
significant at a 13\% (0.17\,$\sigma$) confidence level.

As Walker et al. (2006) discuss, apparent rotation can also originate
from relative transverse motion of a dSph due to the bulk proper
motion of the system. As there are no extant observations of Leo\,II's
proper motion, owing to its large distance, we cannot approach the
question of whether the relative motion between the Sun and Leo\,II would
be sufficient to account for the full amplitude of our observed
rotation.
All in all, there is no kinematical evidence of any significant
velocity gradients in Leo\,II, either due to rotational support or
produced by Galactic tides. It therefore appears that tides most
likely have not affected Leo\,II to any significant degree.

\section{Radial variation of the dispersion}
In order to derive the radial velocity dispersion profile of Leo\,II we
employed a maximum likelihood technique, which is described in more
detail in Kleyna et al. (2004) and Koch et al. (2007a).  In essence,
we radially binned our data such that the same number of stars per bin
was maintained.  The binsize was defined so as to ensure a
sufficiently large number of stars per bin, and we chose to include no
fewer than 12 data points in each bin.  A Gaussian velocity
distribution centered on the single systemic mean velocity is then
assumed for each radial bin. This distribution is convolved with the
observational errors and also with an additional component, which
accounts for the Galactic foreground contamination. The latter is
represented by a uniform interloper velocity distribution contributing
a fraction $f_{int}$ to each individual bin (see eq. 2 in Koch et
al. 2007a).
As the velocities of our non-member stars also appear to be well fit
by a power-law distribution with an exponent of $-0.6$, we also
considered such an interloper distribution in our determinations of
the dispersion profile.  Nevertheless, it turned out that the
resulting velocity dispersion profiles are indistinguishable in
practice and $f_{int}$ is typically compatible with zero in all
bins. Finally, the error bounds on the dispersion were calculated by
numerically integrating the total probability of the data set and
finding the corresponding 68\% confidence intervals.  Fig.~7 displays
the radial dispersion profile thus obtained.

Given the results of Sect.~3, we neglected the effect of a
rotationally-induced inflation of the dispersion in our calculations.
Another critical issue in the proper dynamical study of a stellar
system is the influence of any significant binary population.  The
presence of any such component in kinematical data leads to an
additional, artificial inflation of the observed line of sight
velocity dispersion. Furthermore, the high-velocity extension of a
binary distribution increases the deviation from a Gaussian so that
the assumption of a zero binary fraction yields larger error
estimates, which may invalidate the assessment of any radial
information derived from the dispersion profiles.

However, it has been shown via Monte Carlo simulations in the past
that the influence of binary systems in dSphs is in fact negligible,
as the purely binary induced dispersion tends to be small compared to
the overall large velocity dispersions found in these galaxies
(Hargreaves et al. 1996; Olszewski et al. 1996).  Furthermore, repeat
observations of red giant velocities in the Draco dSph (Kleyna et
al. 2002) and in Fornax (Walker et al. 2006) suggest that the impact
of binaries on the measured velocity dispersion is negligible.  These
kinematical studies did not support an overall binary content larger
than 40\% in those dSphs.  Also, Kleyna et al. (2002) note that the fraction of 
dynamically significant binaries in the Draco sample amounted to less
than 5\%.  Likewise, Koch et al. (2007a) showed for the case of the
Leo\,I dSph that the assumption of a larger  binary fraction does not significantly
alter the resulting velocity dispersion profile.  Finally, there seems
also to be no evidence that binaries would have significantly affected
Leo\,II's internal kinematics, as revealed by the Monte Carlo simulations
of V95.  If we interpret observed velocity differences as an
indication of binarity, there is only one object with a difference in
velocity between our measurement and that of V95 that is in excess of
2$\sigma$ (taken as the measurement uncertainty).  If one was to
interpret this in terms of binarity (not accounting for any selection
effects in the data), this would also suggest a low binary fraction in
Leo\,II. {The lack of repeat measurements of the stars in our data set inhibits  
an assessment of potential binaries.}
Hence, we will proceed by neglecting the presence of any binary
population in our calculations.

The resultant velocity dispersion profile (Fig.~7) is essentially flat
out to the last data point and at most subject to fluctuations within
the measurement errors.  The last point, shown at a median distance of
0.5\,kpc in Fig.~7, includes our outermost observed targets at
locations of 1.5\,kpc from the galaxy's center.  Overall, the profile
is well described by a constant value of 6.6\,\kms, which is in
agreement with Leo\,II's global value derived in Sect.~2.3.
Hence it appears that Leo\,II does in fact show the same behavior as the
majority of the dSphs analysed to date, in terms of an essentially
flat velocity dispersion profile (Wilkinson et al. 2006 and references
therein).  This suggests that, similar to the other dSphs, this galaxy
is also a dark matter dominated system out to the largest scales.  We
will turn to the dynamical implications of this finding by deriving
Leo\,II's mass profile and mass to light ratio in Section~5.
\subsection{Correlation with metallicities}
For 52 of the targets around Leo\,II's velocity peak, metallicities were
derived in \p1.  From these it was inferred that Leo\,II exhibits a wide
range of metallicities, covering at least 1.3\,dex. The derived
age-metallicity relation indicates that Leo\,II experienced constant star
formation over an extended period. Overall, this galaxy appears to
have formed stars from 15\,Gyr until 2\,Gyr ago, although its dominant
stellar population is of intermediate age, around 9\,Gyr (Mighell \&
Rich 1996).
When split into a metal-poor and a metal-rich population by dividing
at the median metallicity, it was shown in \p1 that these stellar
components do not differ significantly in their spatial distributions,
i.e., there is no apparent metallicity gradient present in
Leo\,II. Moreover, the stellar ages do not appear to show spatial trends.
It is then worth noticing that the star formation histories derived from 
deep, but central, pencil beam surveys such as the HST studies of 
Mighell \& Rich (1996) and Dolphin (2002) are also consistent with our findings 
in \p1 from a wide-field survey. This again argues in favor of the lack of any 
considerable population gradient between the center and Leo\,II's outskirts. 
In dwarf galaxies with extended star formation histories, the younger
and/or more metal-rich populations generally tend to be more centrally
concentrated (e.g., Harbeck et al.\ 2003).  
Moreover, different
stellar populations in dSphs may also have significantly different
kinematics, as in the case of Sculptor (Tolstoy et al. 2004) or CVn
(Ibata et al. 2006). 

In our data, there is no apparent trend of velocity with metallicity
discernible (top panel of Fig.~8).  In the following, we also
separated our velocity data according to their previously assigned
metallicities.  The respective radial velocity histogram is shown in
Fig.~8 (left panel).  It turns out that the metal poor stars have
slightly higher velocities on average, but the difference between them
and the metal richer member candidates is marginal. Using the same
formalism as outlined in Section~2.3, we find the mean velocity of the
metal poor stars to be (79.1$\,\pm\,$1.6)\kms\ versus
(76.9$\,\pm\,$1.0)\kms\ for the metal rich component.  The overall
velocity dispersions for the two sample are indistinguishable and are
consistent with that obtained using the entire velocity data set.  In
addition, the data were divided into three radial bins (see Fig.~8,
right panel).  In this case, the metal poor (and presumably older; see
\p1) component has a higher velocity dispersion in each bin.  
However, given the low number of stars per bin (of order ten at our
chosen binning) and the derived measurement uncertainties, any
difference in the dispersion between both metallicity populations is
in fact insignificant.  Thus we conclude that Leo\,II's stellar
populations are kinematically indistinguishable and well-mixed in
phase-space, which is fully consistent with the lack of any obvious
metallicity or age gradient (Bellazzini et al. 2005; \p1).

\section{Mass, light and mass-to-light estimates}
In order to define a mass estimator that can be readily implemented 
 we integrated the 
Jeans equation (Binney \& Tremaine 1998, eqs. 4-54 ff.; Koch et
al. 2007a) under the assumption of spherical symmetry and an isotropic
velocity distribution.  Since our data cover the entire face of the
galaxy, this procedure has the advantage of exploiting the full
kinematic information that has been gathered in the previous sections
as opposed to mass estimates that rely on the single value of a
central velocity dispersion (e.g., V95; Coleman et al. 2007 for the
case of Leo\,II).  Moreover, these methods of ``core-fitting'' make the
assumption that the mass distribution follows that of the light
(Richstone \& Tremaine 1986; V95), which has been shown to be invalid in
the case of the Draco dSph (Kleyna et al. 2002). Nevertheless, a
limitation of the simple formalism employed in this section is the
neglect of the degeneracy between velocity dispersion and anisotropy
(Wilkinson et al. 2002), which persists unless large data sets and/or
higher moments of the velocity distribution are considered (e.g., 
Binney \& Mamon 1982; \L okas \& Mamon 2003).  
We will account for this effect in detail in Sect.~6.

The surface brightness profile required for the mass determinations
was adopted from a fit to the data of Irwin \& Hatzidimitriou (1995). It is 
well fit by a Plummer profile with a characteristic radius of
$2.76\arcmin\pm 0.09\arcmin$ (see Fig.~9), where the error was
determined via bootstrap resampling (e.g., Mackey \& Gilmore 2003).
This value is in fact very similar to Leo\,II's core radius from a King
(1966) profile fit (Irwin \& Hatzidimitriou 1995; Coleman et al. 2007)
and use of the latter value yields analogous results for the mass
computations.
As implied by the results from the previous sections the dispersion
profile was assumed to be radially constant.

The resultant mass profile of Leo\,II is shown in Fig.~9. From this we
estimate its total mass out to the tidal radius at 8.7$\arcmin$
(0.6\,kpc) to be $(2.7\,\pm\,0.5)\times10^7$\,M$_{\odot}$. The
quoted errors were obtained from a Monte Carlo simulation of the mass
calculation and take into account the 1$\sigma$ confidence interval 
of the velocity dispersion as well as the
uncertainty in the parameters of the light profile.

Also shown in Fig.~9 is the density profile that we obtained from the
above dynamical considerations. It is characterised by a central
density of
$\rho_0=(3.4\,\pm\,0.4)\times10^8\,M_{\odot}$\,kpc$^{-3}$, which is
in good agreement with the value of
$(4.0\,\pm\,1.1)\times10^8\,M_{\odot}$\,kpc$^{-3}$ derived by V95
from their analysis of Leo\,II's very center.\footnote{Coleman et
al. (2007) yield a lower value of
$\rho_0=(1.3\,\pm\,0.4)\times10^8\,M_{\odot}$\,kpc$^{-3}$ under the
assumption of a Gaussian mass and light distribution.}
Leo\,II's density distribution resembles a cored profile in the innermost regions, 
although we note that our innermost target star lies at a distance of
0.5$\arcmin$ (0.03\,kpc) from the center so that any assessment of the
``central'' density has to rely on an extrapolation.  
The profile reaches a power-law slope of $-1$ at approximately
0.13\,kpc and converges towards a slope of $-2$ for distances larger
than 0.5\,kpc.

With these mass estimates at hand, we derive the satellite's  M/L ratio
by adopting its luminosity from the literature. Thence the (M/L) (in
solar units) varies from 26.6$^{+5.5}_{-4.5}$ for
L$_V$\,=\,(9.9$\pm$0.2)$\times$10$^5$\,L$_{\odot}$ (V95),
35.6$^{+8.8}_{-7.7}$
(L$_V$\,=\,(7.4$\pm$2.0)$\times$10$^5$\,L$_{\odot}$; Coleman et
al. 2007) to 45.4$^{+11.7}_{-10.4}$ for
L$_V$\,=\,(5.8$\pm$1.8)$\times$10$^5$\,L$_{\odot}$ (Mateo 1998)\footnote{For the subsequent 
discussions of Leo\,II's M/L, e.g., Gilmore et al. (2007), we adopt the latter value using Mateo's (1998) 
luminosity estimate, since this value is, in turn, based on the measurements of Irwin \& Hatzidimitriou 
(1995), whose surface brightness profile we employed.}.  In all cases, 
a M/L of this order of magnitude strongly suggests that Leo\,II is
governed by a considerable amount of dark matter on all spatial
scales.  
{Mateo et al. (1998) estimate that the {\em stellar} (M/L)$_V$ for a galaxy like Leo\,II with 
prominent old and intermediate age populations (see also \p1) is of the order of 1--1.4. 
If, on the other hand, Leo\,II contained only an old, single-age globular cluster-like  stellar population, 
then Mateo et al. (1998) indicate that the M/L ratio of Leo\,II 
would be higher than the value of the galaxy's actual population by a factor of $\sim$1.3.
 Clearly, these results and our observations show that the dark matter content of Leo\,II is 
well above that expected in a purely old stellar population.} 
From their photometric structural analysis of Leo\,II, Coleman et
al. (2007) derive its M/L via two assumptions: for the case of mass
follows light they end up with a (M/L)$_V$ of 7, whereas for the other
extreme of a pure dark matter dominance, this value may be as high as
$\sim$125.  The fact that our full kinematical analysis 
points towards the higher end of these directions underscores the dark
matter dominated state of Leo\,II, with at most mild tidal perturbation.
{Since the spatial resolution of our data and the resulting method 
is an improvement over the traditional core-fitting methods, which assume a constant 
M/L, we show in Fig.~9 (bottom panel) also the theoretical M/L profile over the full range 
covered by our observations and using 
the previously derived parameters. Already at Leo\,II's center the M/L is an order of magnitude higher than 
the purely stellar estimates.} 

Given the above results, Leo\,II fits well remarkably well on the (M/L)
versus $M_V$ relation for dSphs (see Fig.~5 in Gilmore et al. 2007),
which is an oft-used argument in favor of all dSphs being embedded in
dark halos of the same mass. Although the exact value of such a common
mass scale is subject to various assumptions such as the M/L of the
pure stellar component, this underlying halo tends to be of the order
of 3$\times$10$^7$\,M$_{\odot}$ (Mateo et al. 1993; Wilkinson et
al. 2006; Gilmore et al. 2007).

\section{Velocity Anisotropy -- cusp versus core}

All the mass computations in Sect.~5 relied on the assumption of an
isotropic velocity tensor, i.e., the anisotropy parameter
$\beta=1-{\left<v^2_{\theta}\right>}/{\left<v^2_r\right>}$ was assumed
to be zero.  Nonetheless, there is compelling evidence that the shapes
and kinematics of most dSphs necessitate non-negligible amounts of
velocity anisotropy on all spatial scales (e.g., \L okas 2002; Kleyna
et al. 2002; Wilkinson et al. 2004; Klimentowski et al. 2007). As the
neglect of a non-isotropic velocity distribution can considerably
affect estimates of the underlying dark halo mass, we explore in the
following the possible parameter space of Leo\,II's velocity anisotropy.
In doing this we follow the same formalism as outlined in Koch et
al. (2007a) (see also \L okas 2002).  That is, under the assumption of
a (constant) anisotropy parameter and for different plausible mass
profiles, we numerically determined a theoretical velocity dispersion
profile; the model parameters were then optimized in a least-squares
sense to yield the best representation of the observations from
Sect.~4.

\subsection{Density cusps} 
Motivated by the cosmological simulations of Navarro, Frenk \& White
(1995; NFW), the density profiles of dark matter halos are often
described by a convenient function with only one free parameter, the
``characteristic density''. For simulated haloes, this
parameterisation has proven applicable to a wide range of halo masses
from the smallest scales, such as the dSphs' dark matter associations,
up to the large scale structures of galaxy clusters.  We first adopted
this cuspy halo mass model,
\begin{equation}
M(r)\,\propto\,M_v\,\left\{\ln (1+cr/r_v)\,-\,\frac{cr/r_v}{1+cr/r_v}\right\}, 
\end{equation}
in our computations of an empirical dispersion profile. 
In our representation, the only free parameter is the virial halo mass, $M_v $, 
and all other parameters, namely virial radius $r_v$ and concentration $c$, are
empirically scaled with this mass according to the relations obtained
by Jing \& Soto (2000) (see \L okas \& Mamon 2001; eqs. 9--12 in Koch et
al. 2007a for details of the parameterisations).

The best-fit results for varying degrees of anisotropy are overlaid on
the observed dispersion profiles in Fig.~10. For all choices of
$\beta$, the virial mass of the NFW-halo is of the order
2.3$\times$10$^8$M$_{\odot}$, corresponding to a formal virial radius
of about 12\,kpc (which is much 
larger than the stellar limiting radius, allowing for the factor of $\sim$10 in  mass out to the 
virial radius).  As for the anisotropy parameter, the reduced
$\chi^2$ for a $\beta$ of ($-1$, $-$0.5, 0, 0.5) amounts to (0.29,
0.21, 0.22, 0.47).  The best coincidence of prediction and observation
is obtained when a slight amount of tangential anisotropy
($\beta\sim-0.25$) is included.

\subsection{Cored density profile}
The majority of presently available observational data of
low-luminosity galaxies indicates that their kinematics and structures
are consistent with cored density profiles (e.g., \L okas 2002; Kleyna
et al. 2003; Salucci et al. 2003; S\'anchez-Salcedo et al. 2006).
Hence, we also computed dispersion profiles in the same manner as
above for a halo profile with variable inner logarithmic slope
(Hernquist 1990; Read \& Gilmore 2005; eq. 13 in Koch et al. 2007a): 
\begin{equation}
\rho(r)\,\propto\,C\,\left(\frac{r}{r_s}\right)^{-\gamma}\,\left[1+\left(\frac{r}{r_s}\right)^{\alpha}\,
\right]^{(\gamma-3)/\alpha}
\end{equation}
The parameters were then determined such as to match the observed dispersion 
profile for different values of the anisotropy $\beta$. 

As it turns out, the ``best-fit'' for all choices of anisotropy was obtained for a slope parameter
of $\gamma\simeq0$, which corresponds to a pure core, as opposed to
the cuspy halo class ($\gamma=1$).  Moreover, a scale radius $r_s$  of
0.3\,kpc and a large smoothness parameter $\alpha$ (which determines
the sharpness of the transition toward the density profile at large
radii) of $\sim$3.7 were found to yield the best representation of our
data (see lower panels of Fig.~10).  For this cored model, the values
of the reduced $\chi^2$ are (0.21, 0.20, 0.21, 0.31) for constant
anisotropies of ($-1$, $-$0.5, 0, 0.5) and, as for the cuspy halos,
our data ideally require some tangential anisotropy of the same order
of magnitude, that is $\beta\sim-0.25$.

It is worth noticing that our data do not allow us to unequivocally
constrain the shape of the velocity anisotropy tensor at large radii
for either choice of halo profile.  However, we can conclude from
Fig.~10 that strongly radial anisotropy is not suitable to account for
the dispersion profile towards the central regions.

\subsection{Cusp or Core ?}
For both profiles investigated above, the mass at the galaxy's
limiting radius amounts to $\sim2.0\times10^7$M$_{\odot}$ and is
thus slightly smaller than that derived from a purely isotropic
velocity distribution in Sect.~5\footnote{The assumption 
of velocity isotropy would generally lead to 
an overestimate of the mass at large radii, if Leo\,II were indeed tangentially 
anisotropic, whereas  if it were radially anisotropic in the outer parts, 
we would underestimate the mass assuming isotropy.}.

Judging by the values of the $\chi^2$ statistics, a cored density
profile is favored to account for the  Leo\,II observed kinematics; however,
the difference between this and the cuspy model is small.  In
particular, there is an indication of a leveling off of the
predictions for the dispersion towards larger radii for the cored model, a trend that
merits further investigation from a large sample of stars at and
beyond the limiting radius of this galaxy.

In addition, we re-ran the computations with a velocity anisotropy
which varies radially according to the prescription of Osipkov (1979)
and Merritt (1985), i.e., $\beta(r)=r^2/(r^2+r_a^2)$, which yields radial 
anisotropy for the case of  $r>r_a$.  
The resulting
curves are shown in the right panels of Fig.~10.  While the NFW halo
requires an anisotropy radius $r_a$ of $\sim$0.7\,kpc and thus larger
than Leo\,II's tidal radius, the cored profile is best associated with a
$r_a$ of $\sim$0.2\,kpc. The respective $\chi^2$ statistics are then
0.22 and 0.19, thus again marginally favoring a cored halo\footnote{We note 
that a radially varying tangential anisotropy $\beta(r)=r^2/(r^2-r_a^2)$ (Merritt 1985; 
his model II) yields essentially the same results.}

As a comparison with extant data of the density profiles of Local
Group dSphs indicates, the majority of these systems are consistent
with cored inner halo profiles, which is similar to our findings for
Leo\,II (Wilkinson et al. 2006; Gilmore et al. 2007 -- see their
Fig.~4). While the outer density profile of Leo\,II is similar to those of
the other LG dSphs analysed to date,
Leo\,II  apparently exhibits a mildly steeper central density profile than 
the other dwarfs, due to the apparently smaller size of its central
core.

From their photometric analysis Coleman et al. (2007) argue that Leo\,II's
M/L ratio has an upper limit of $125^{+56}_{-51}$ under the assumption
of a constant density core out to their tidal radius. Although our value derived in the previous
Section is lower by a factor of approximately 3, we note that the core
size we obtain in the density profile is correspondingly smaller.

\section{Summary} 
We have obtained a large data set of radial velocity measurements for
about 171 red giant member candidates in the remote Leo\,II dSph, thereby
increasing existing published samples more than fivefold (V95).  These
data extend in radial distance out to the galaxy's limiting radius,
which is an important prerequisite for kinematic studies, since Leo\,II,
as the second most remote dSph satellite of the Milky Way, is valuable
for an assessment of the potential role of Galactic tides.  With its
moderate present-day systemic velocity, Leo\,II can be considered as bound
to the Galaxy and yet there is no indication of any tidal
perturbation. This lack of perturbations is expected if this dSph has
maintained its present large distance over long time-scales.  Our
study precludes any apparent rotation of Leo\,II's stellar component;
moreover we do not detect a velocity gradient, nor any asymmetry in
the sense that high and low velocity outliers preferentially reside in
particular directions along the major axis, which might reflect the
action of tides.  Although other dSphs may well exhibit ``extratidal'' 
material (e.g., Carina; Mu\~noz et al. 2006) or be in an advanced
state of dissolution (e.g., Sagittarius; Ibata et al. 1994), these
effects are clearly not seen in Leo\,II.
Thus we must conclude that this galaxy is a purely pressure-supported
system which has not been significantly affected by tides over the
course of its evolution.

The radial velocity dispersion profile and resulting density profiles
show a resemblance to the respective profiles of the majority of the
other Local Group dSphs, and display the well-established features of
these systems, namely a dispersion which is flat out to the reach of
the present data and a density profile that is consistent with a cored
halo mass distribution. We note, however, that our present data do not
unambiguously distinguish between the possibilities of a cusped or
cored density profile in terms of the $\chi^2$-statistics of the
best-fit to our observed dispersion profile.  Both cored and cusped
halo models favor the presence of a mild amount of tangential
velocity anisotropy in the central regions of the galaxy.

All in all, we find that Leo\,II is a typical dSph in terms of the values
of its total mass of $\sim$3$\times$10$^7$\,M$_{\odot}$ and of its
characteristic central density of the order of
$3.4\times10^8\,M_{\odot}$\,kpc$^{-3}$ (Gilmore et al. 2007).
Depending on the adopted total luminosity of Leo\,II, we estimate its M/L
ratio to lie in the range 25 to 50. This is, in conjunction with the
flatness of the dispersion profile, striking evidence that Leo\,II is a
dark matter dominated system.  Moreover, the value we find is
consistent with the idea that all dSphs are embedded in a dark matter
halo with one single characteristic mass scale.  Taking into account
that all nearby dSph galaxies contain very old stellar populations
(Grebel \& Gallagher 2004), it thus appears reasonable to consider
these objects as the smallest surviving condensations of dark matter
with associated stellar populations in the universe (Read et
al. 2006b).

\acknowledgments

A.K. and E.K.G. are grateful for support by the Swiss National Science
Foundation through grants 200020-105260 and 200020-113697.
M.I.W. acknowledges the Particle Physics and Astronomy Research
Council and the Royal Society for financial support.

\clearpage
\begin{figure}
\epsscale{0.8}
\plotone{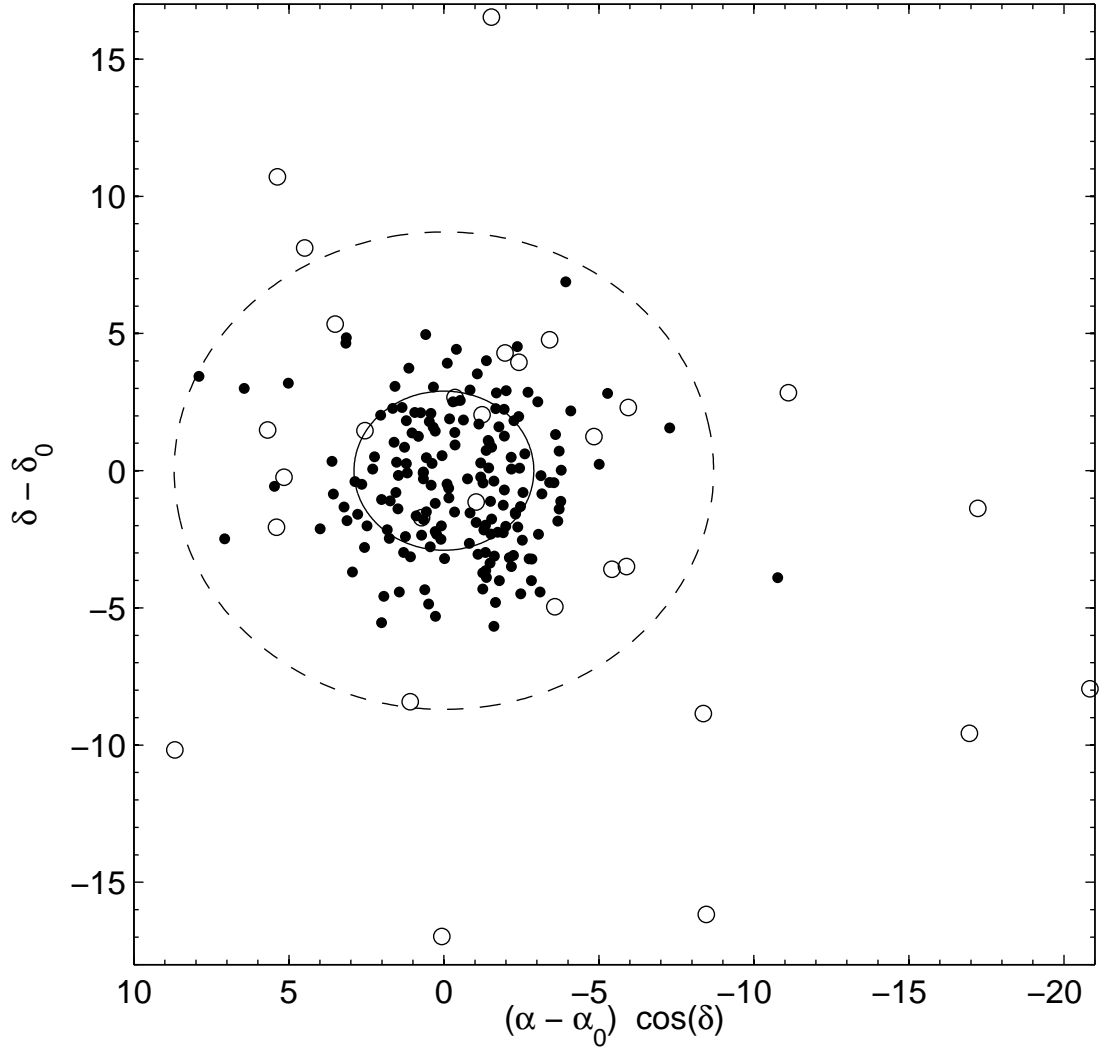}
\caption{Distribution of our targets on the sky, centered on Leo\,II. 
Apparent radial velocity non-members are designated 
with open circles. The solid and dashed lines indicate Leo\,II's core and nominal King tidal radius 
at 2$\farcm$9 and 8$\farcm$7 (Irwin \& Hatzidimitriou 1995).}
\end{figure}
\begin{figure}
\epsscale{1.0}
\plotone{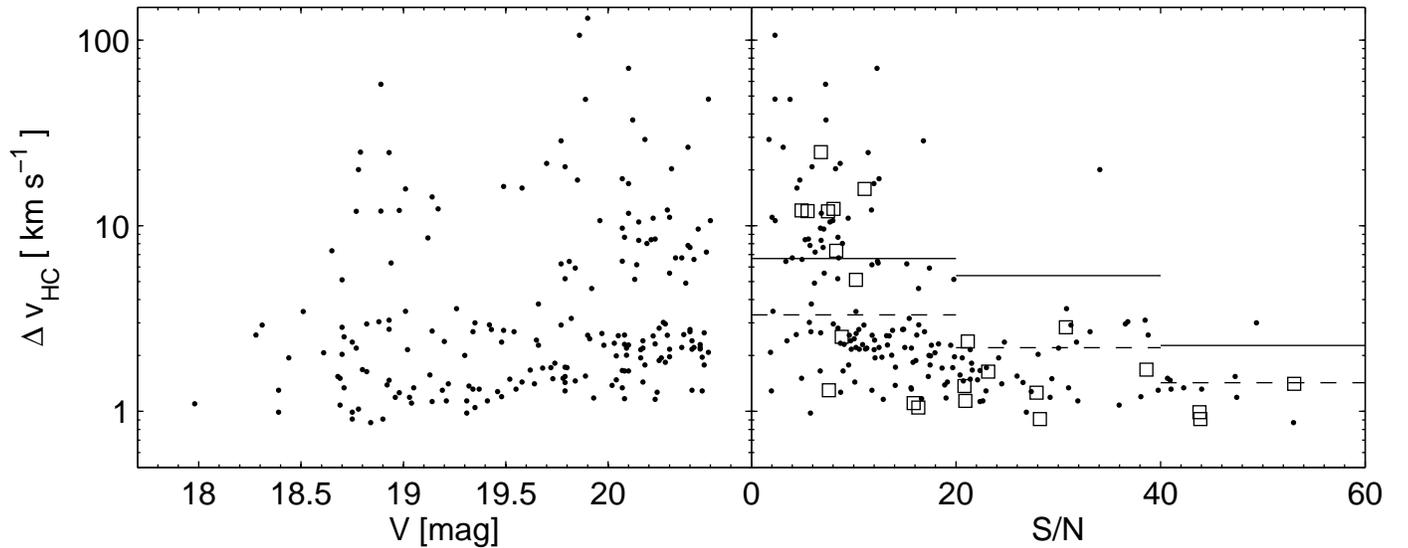}
\caption{Velocity errors versus V-band magnitude (left panel) and {Signal to Noise ratio per pixel 
(right panel). The dashed lines in the right panel indicate our formal mean velocity errors in each S/N 
bin, whereas the solid lines show the r.m.s. discrepancy between our measurements and those of V95 
(open squares), 
which we use to re-scale our uncertainties in Sect.~2.4.}}
\end{figure}
\begin{figure}
\epsscale{1.0}
\plotone{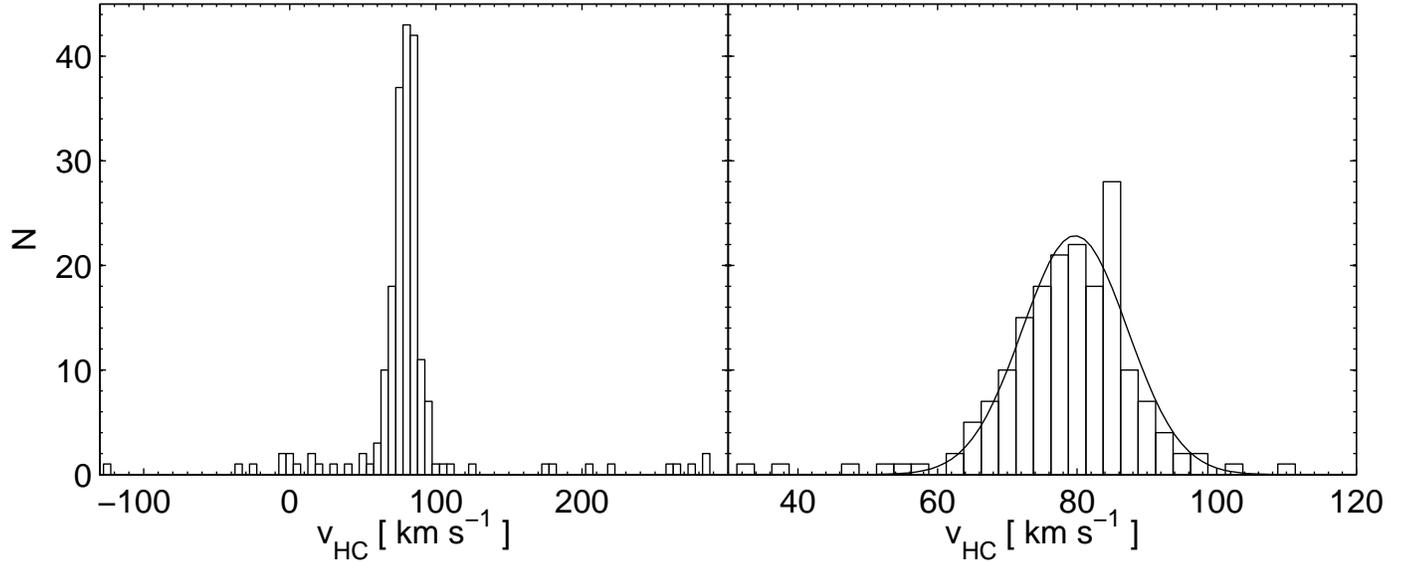}
\caption{Histograms of the radial velocities of stars in our survey of Leo\,~II. 
The left panel displays the full sample, whereas the right panel only
shows stars around Leo\,II's systemic velocity. The best-fit Gaussian to
this distribution is shown as a solid line.}
\end{figure}
\begin{figure}
\epsscale{0.6}
\plotone{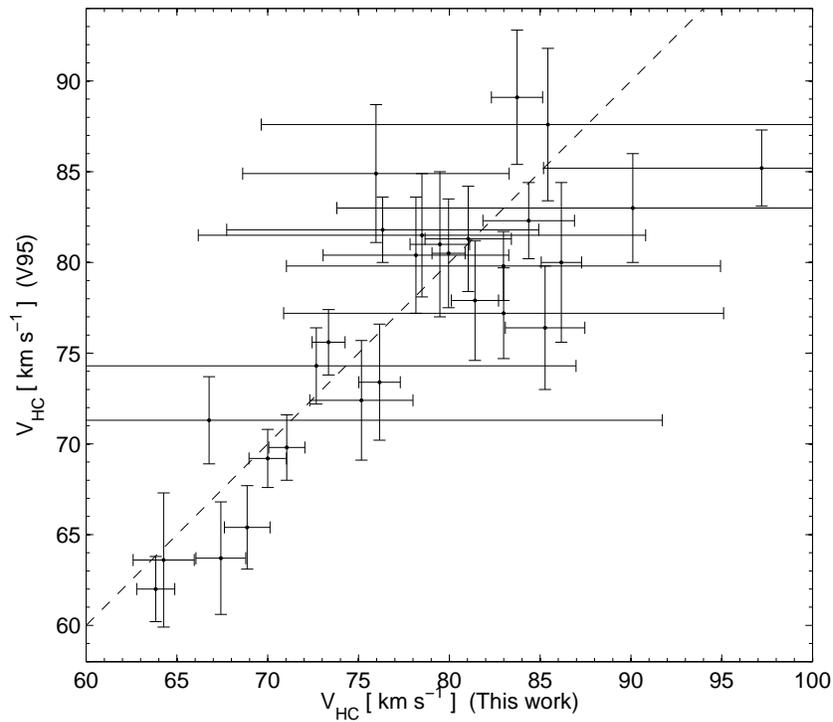}
\caption{Comparison of our velocity measurements to those
estimated by Vogt et al. (1995) for the 28 stars in common between the
two data sets. {The scale factors to the error bars have not been applied in this 
diagram to illustrate our internal formal measurement errors}. The dashed line is unity.}
\end{figure}
\begin{figure}
\epsscale{0.8}
\plotone{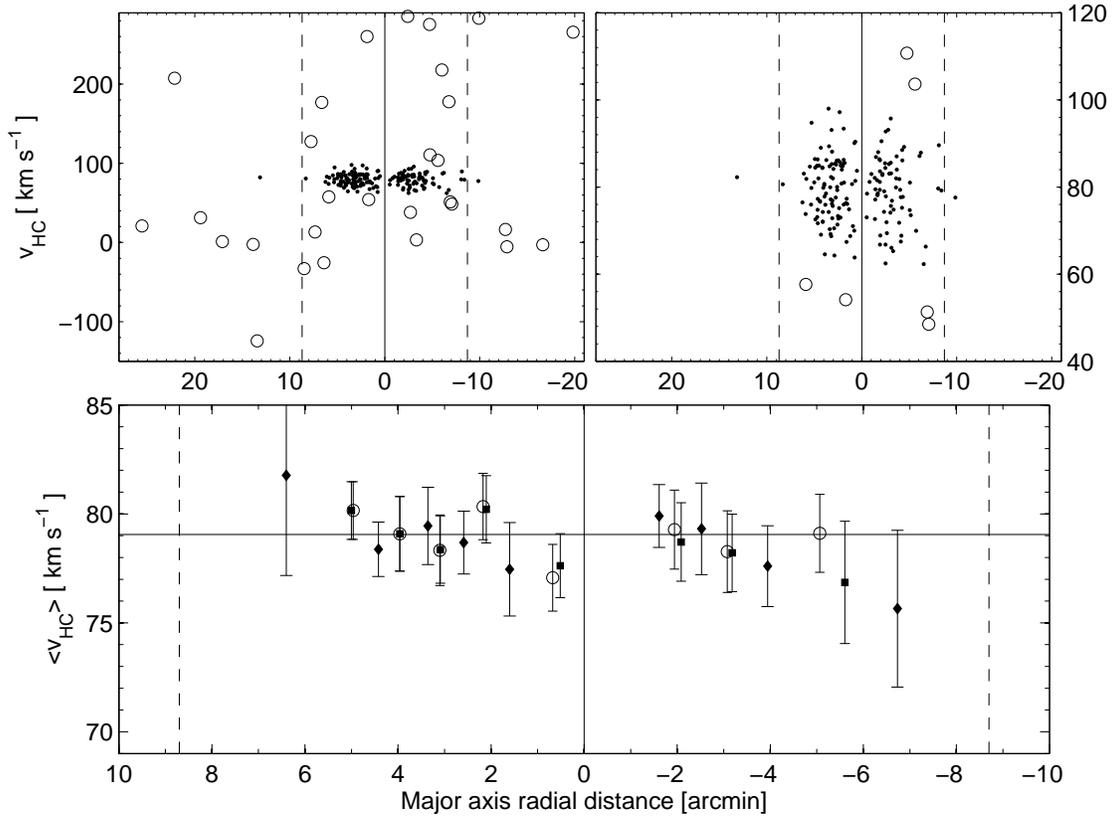}
\caption{Top panels: Radial velocity versus major axis radial distance. 
The right panel shows those stars around Leo\,II's mean systemic velocity, while 
the full data are included in the left panel. Members within (outside) a 3\,$\sigma$ 
cut are shown as filled (open) symbols. 
The bottom panel
displays mean radial velocities obtained in radial bins with a constant number of stars, where we 
adopted different cuts in velocity to consider the presence of high- or low-velocity member stars: 
3\,$\sigma$ (open circles), 5\,$\sigma$ (solid squares), and 10\,$\sigma$ (filled diamonds). 
No significant radial gradient is discernible in our data. {Monte Carlo tests  show 
that the hint of a change 
in the mean velocity between the outer bins at each end of the major axis 
is not statistically significant (see text for details).} 
Note the different scales in each of the subplots.}
\end{figure}
\begin{figure}
\epsscale{0.85}
\plotone{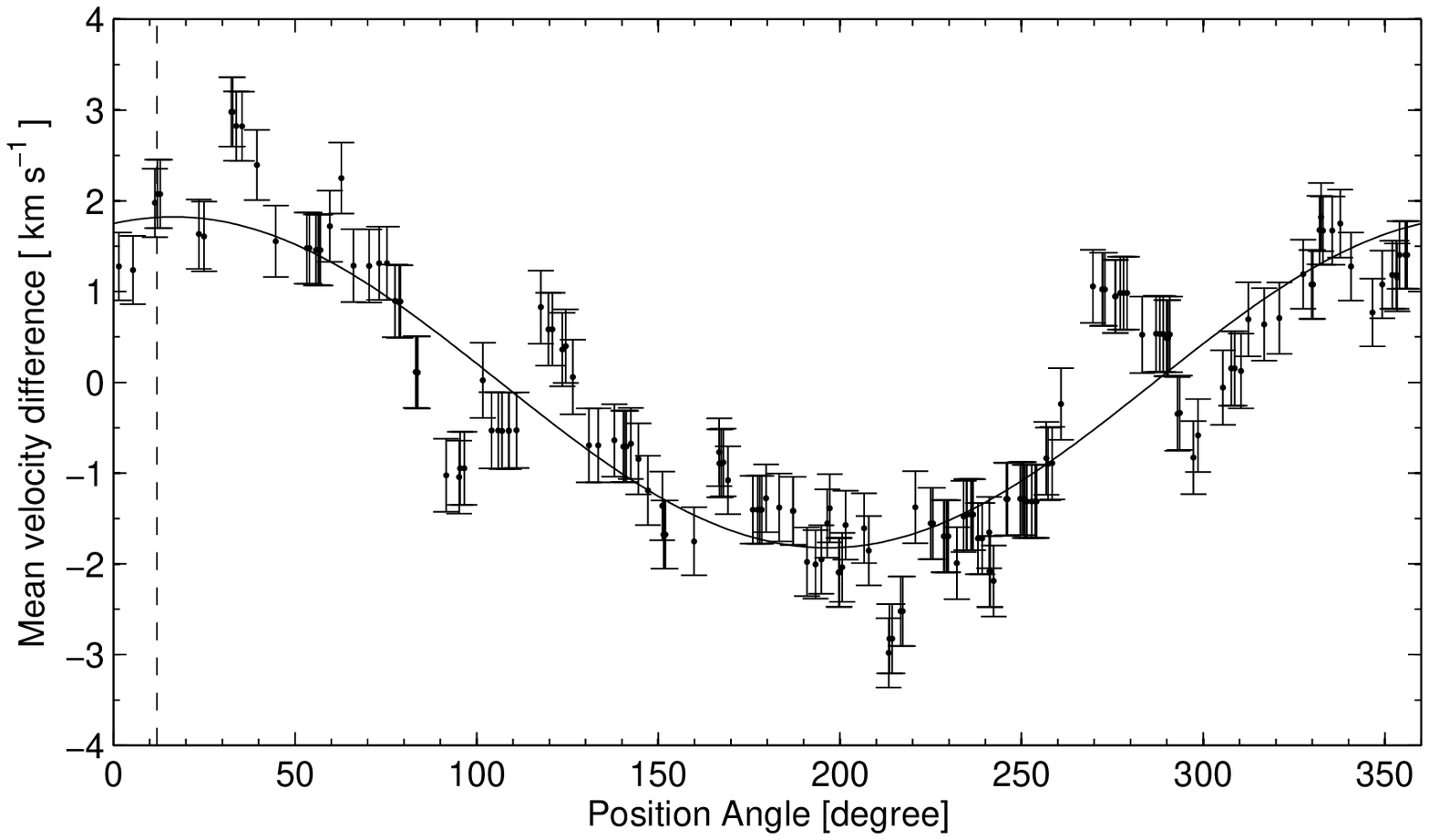}\plotone{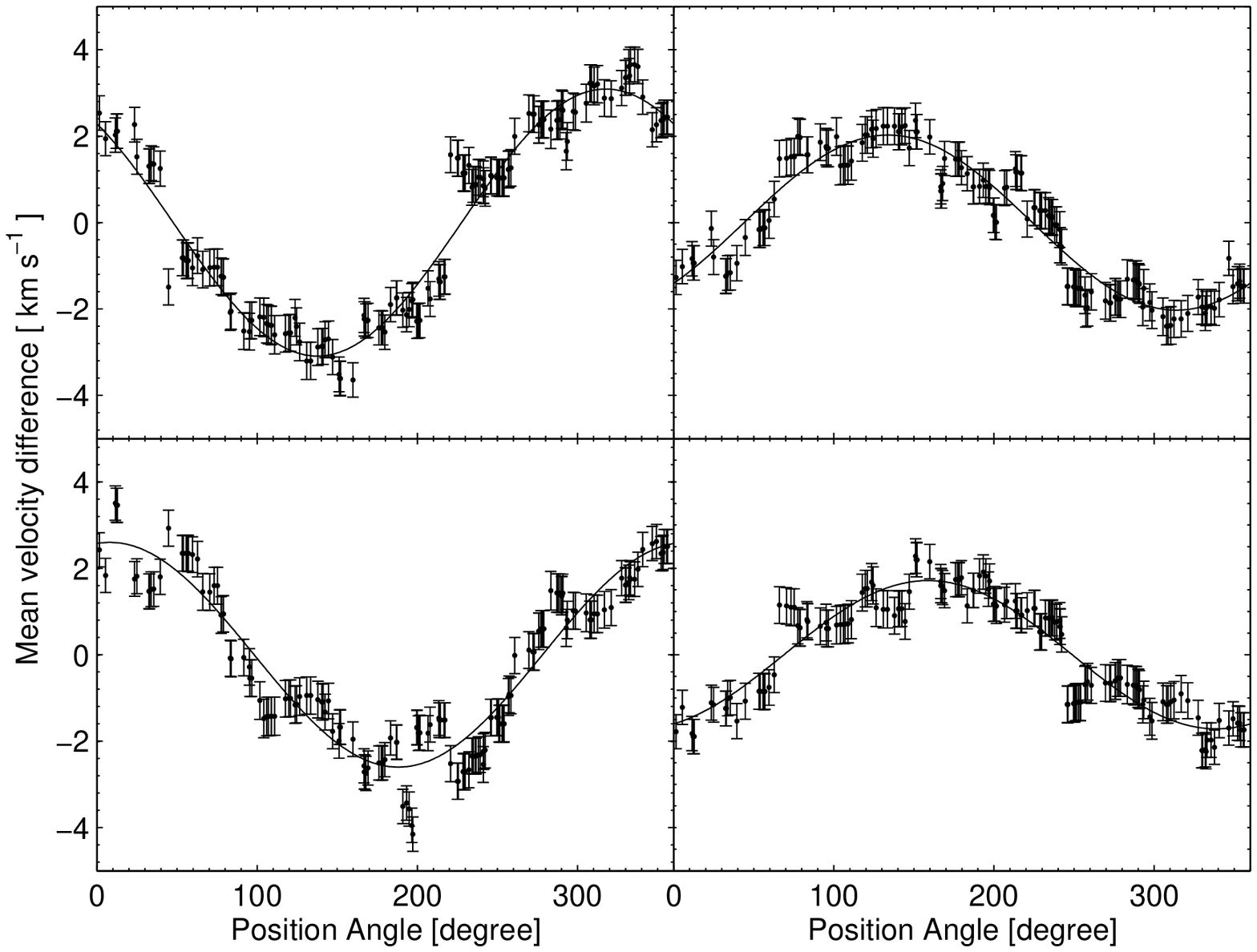}
\caption{Test for apparent rotation {in our observations (top panel) and a sample of 
Monte Carlo simulations (bottom panel).} 
The points with error bars show the mean velocity difference 
of stars on either side of bisecting lines at the respective position
angles.  The best-fit sinusoid is displayed as a solid line. The
discernible rotation amplitude at a position angle of $\sim$23$\degr$
is statistically insignificant -- 87\% of the random samples as shown in the bottom panel 
produce a comparable amplitude or larger.  Indicated
as a dashed line in the top panel is the minor axis position angle of 12$\degr$ from
Irwin \& Hatzidimitriou (1995).}
\end{figure}
\begin{figure}
\plotone{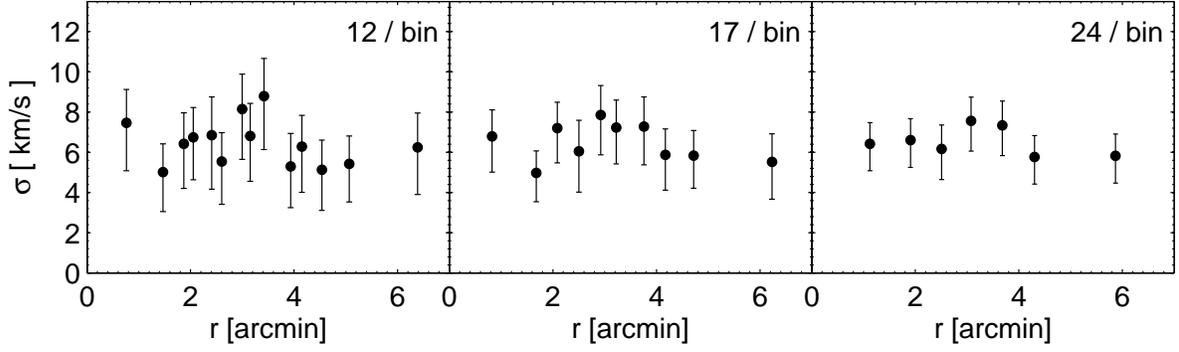}
\caption{Leo\,II's radial velocity dispersion profile obtained for a 
3\,$\sigma$ membership criterion. In each panel, each bin contains the
indicated constant number of stars.}
\end{figure}
\begin{figure}
\epsscale{1.0}
\plotone{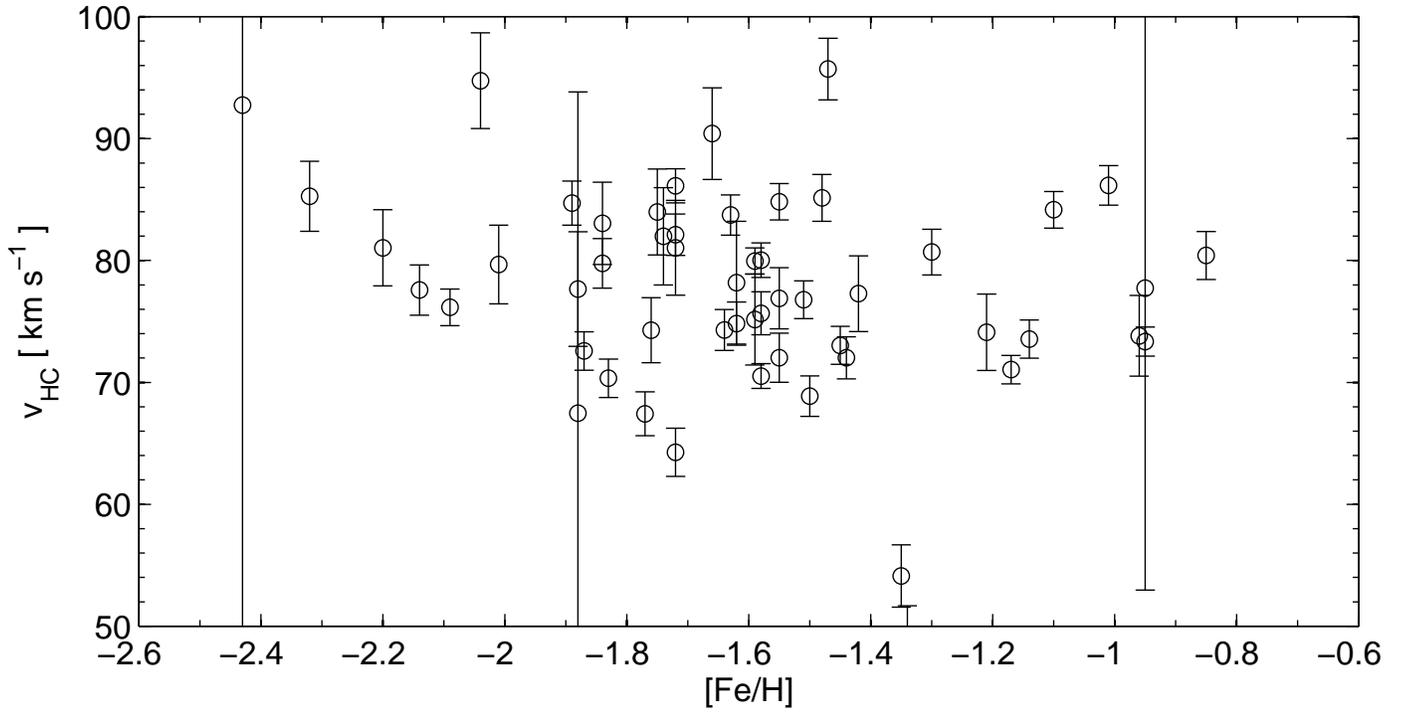}
\plottwo{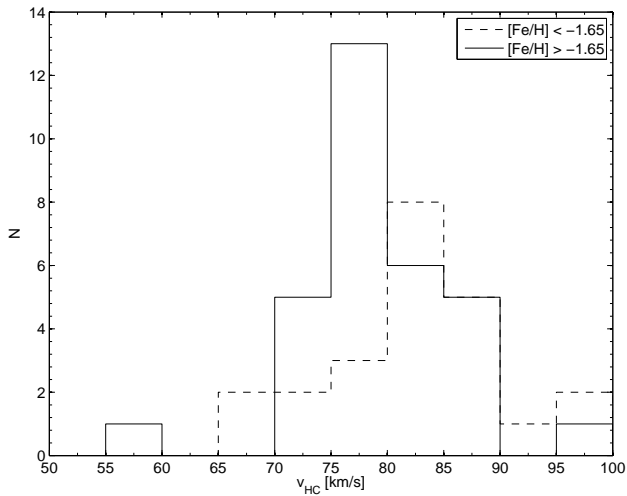}{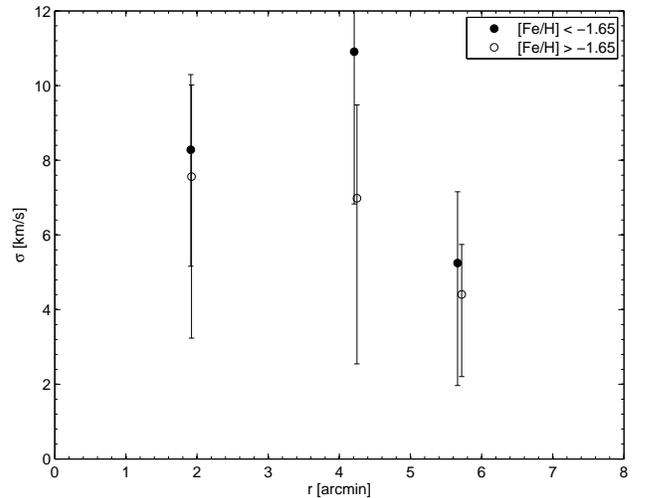}
\caption{Upper panel: Velocity distribution for a subsample of our data for which there is 
extant metallicity information (\p1). The velocity histograms and
dispersion profiles in the lower panels have been separated into a
metal-poor and a metal-rich component, which exhibit only marginally
different kinematics.}
\end{figure}
\begin{figure}
\epsscale{0.8}
\plotone{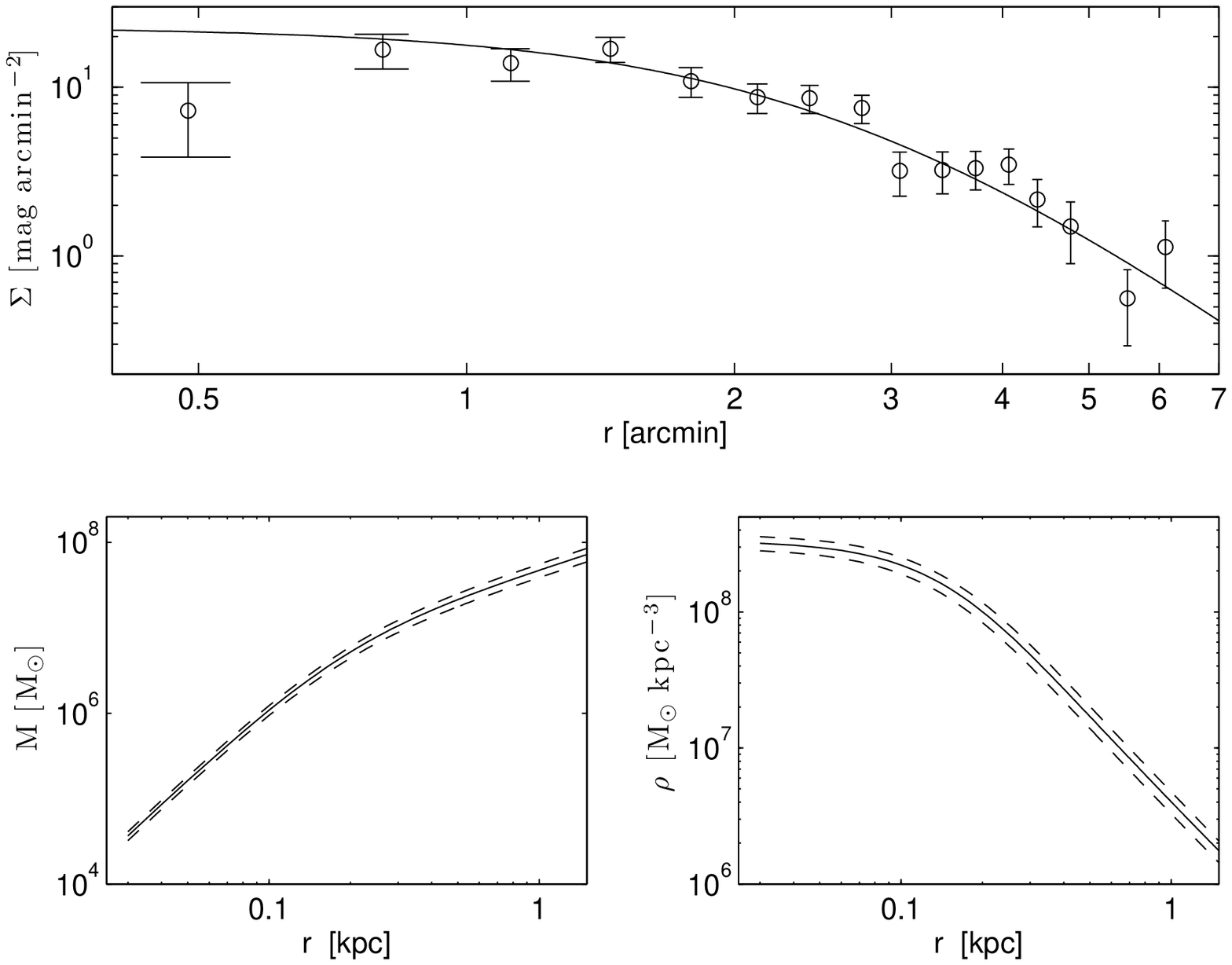}\plotone{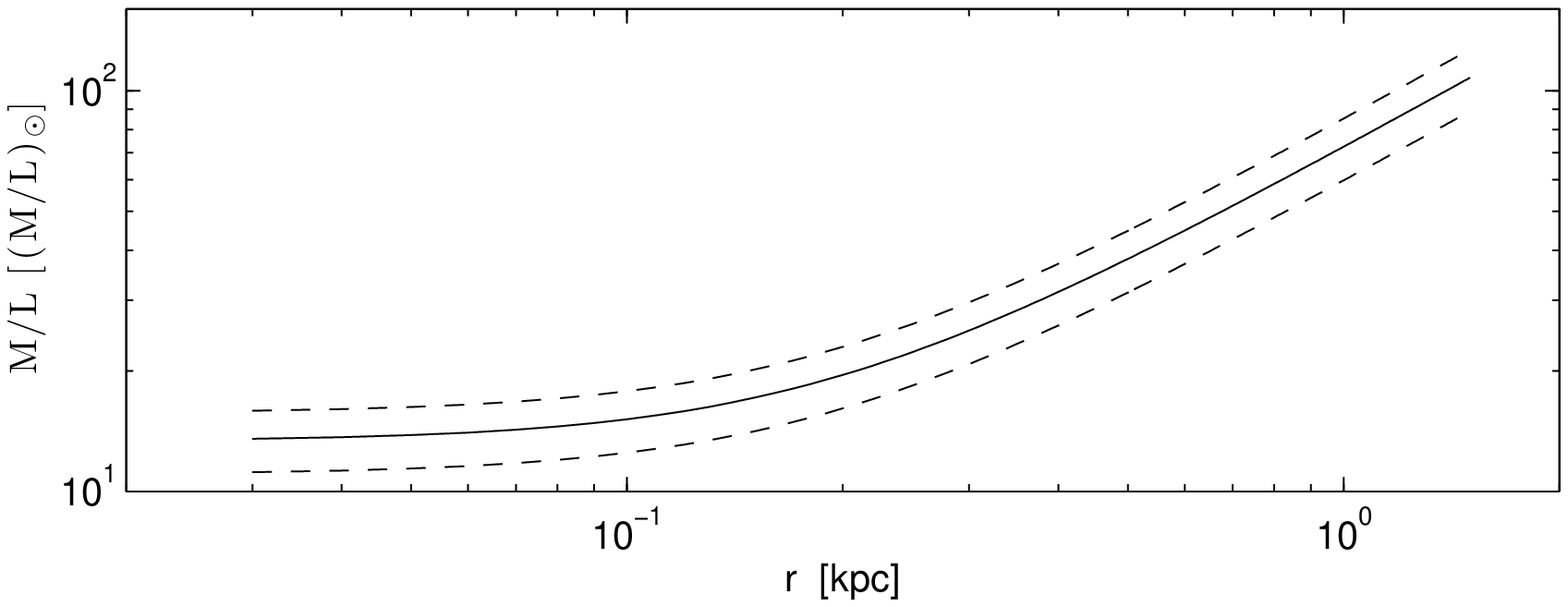}
\caption{Mass (middle left panel) and density (middle right) profile 
of Leo\,II, obtained from Jeans equation under the assumption of an isotropic velocity distribution 
and a constant velocity dispersion. Error bounds originate from the uncertainties in 
the light and velocity dispersion profile. 
The surface brightness profile entering the calculations 
is shown in the top panel (after Irwin \& Hatzidimitriou 1995). The solid line is a best-fit 
Plummer profile. At Leo\,II's distance, 1$\arcmin$ corresponds to about 0.07\,kpc. 
{Based on mass and light profiles, we derive the M/L distribution shown in the lower panel}.}
\end{figure}
\begin{figure}
\epsscale{0.8}
\plotone{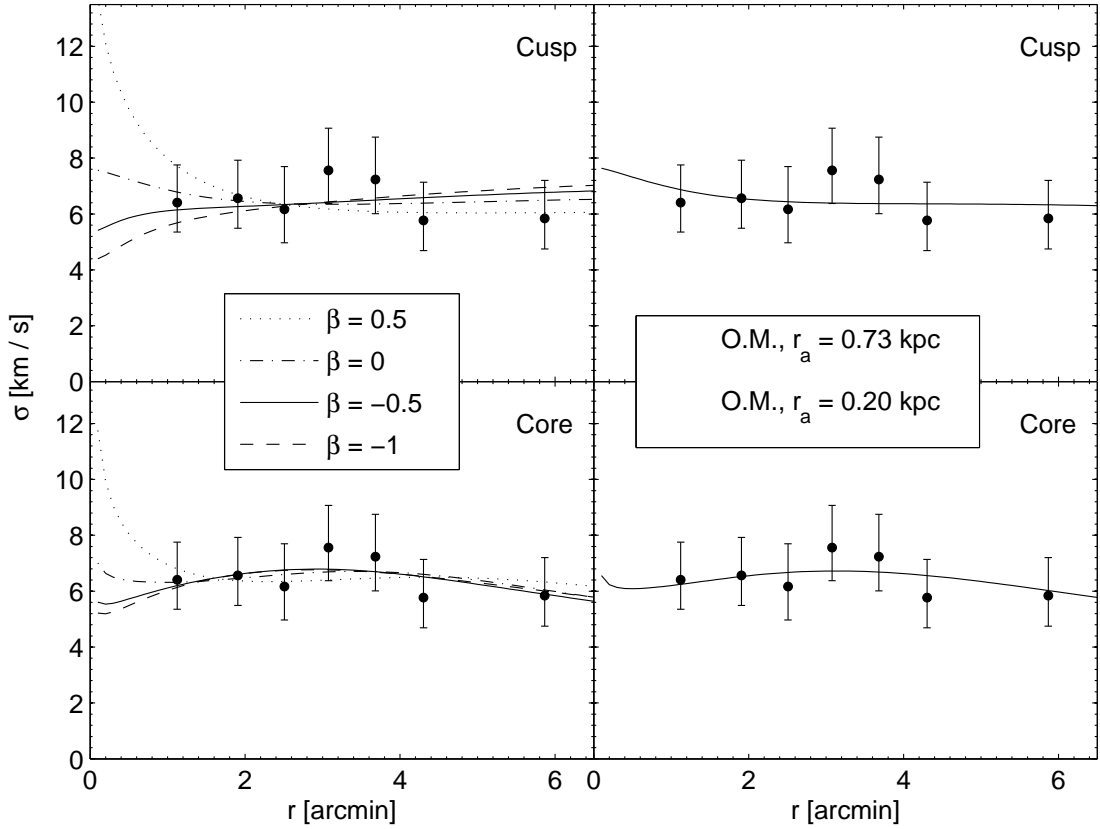}
\caption{Dispersion profiles derived 
under the assumption of varying degrees of velocity anisotropy
$\beta$.  The upper and lower panels adopt two different density
models, namely a NFW cusp (upper panels) and a cored halo-profile
(lower panels), respectively. While the left panels maintain constant
values of $\beta$, the right panels were determined for a velocity
anisotropy which varies radially according to the Osipkov-Merritt
(O.M.) prescription with an anisotropy radius as stated. }
\end{figure}
\clearpage
\begin{table}
\begin{center}
\caption{Measured properties of radial velocity members in Leo\,II}
\begin{footnotesize}
\begin{tabular}{lccrrrcc}
\hline
\hline
Star & $\alpha$ (J2000) & $\delta$ (J2000) & r\,[$\arcmin$] & v$_{\rm r}$ & 
 $\sigma_{\rm v_r}$  & [Fe/H]  & $\sigma_[Fe/H]$ \\
\hline 
T\_18  &  11 13  03.5  & 22 11  35.0  &   7.06 &  48.49  &    3.18  &   $-1.34$  &  0.18\\
T\_19  &  11 13  23.0  & 22 09  23.0  &   1.64 &  78.18  &    5.04  &  $-$1.62 &   0.52\\
V\_19  &  11 13  35.6  & 22 09   06.0  &   1.68 &  76.33  &    8.59  &    \nodata &    \nodata \\
T\_20  &  11 13  32.1  & 22 09  12.0  &   0.77 &  90.10  &   16.29  &    \nodata &    \nodata \\
T\_21  &  11 13  22.7  & 22 08   09.0  &   2.11 &  83.99  &    3.53  &  $-$1.75 &   0.25 \\
\hline
\end{tabular}
\end{footnotesize}
\end{center}
{\footnotesize Note. --- This Table is published in its entirety in the electronic edition of the {\it 
Astronomical Journal}. 
A portion is shown here for guidance regarding its form and content. Metallicities are those from \p1.}
\end{table}

\end{document}